\documentclass[10pt,leqno]{amsart}

\usepackage[utf8]{inputenc}
\usepackage[T1]{fontenc}
\usepackage{graphicx}
\usepackage{multirow}
\usepackage{indentfirst,csquotes}
\usepackage[numbers, sort&compress]{natbib}
\usepackage{amssymb,amsthm,amsmath}
\usepackage{paralist,fancyhdr,etoolbox,xcolor}

\usepackage[colorlinks=true,
  linkcolor=black,
  filecolor=black,
  urlcolor=black,
  pdfencoding=auto,
  psdextra
]{hyperref}

\makeatletter
\renewcommand\section{\@startsection{section}{1}
  {\z@}{2.5ex plus 1ex minus .2ex}{1.3ex plus .2ex}
  {\centering\normalfont\Large\bfseries}}
\renewcommand\subsection{\@startsection{subsection}{2}
  {\z@}{2.25ex plus 1ex minus .2ex}{1ex plus .2ex}
  {\normalfont\large\bfseries}}
\renewcommand\subsubsection{\@startsection{subsubsection}{3}
  {\z@}{2ex plus .8ex minus .2ex}{0.8ex plus .2ex}
  {\normalfont\normalsize\bfseries}}
\makeatother



\begin{document}

\title[Investigation of Double-Pair Anti-Helmholtz Coils]{Investigation of the Benefits and Disadvantages of Using \\ Double-Pair Anti-Helmholtz Coils In BEC-Producing \\ MOT Setups and Optimizing their Design}

\author{Şenol Tarhan}
\author{Gabriel Goetten de Lima}

\address{Address}
\email{talhansenol9@gmail.com}
\email{gabriel.goetten@gmail.com}

\date{\today}

\begin{abstract}
This work has investigated the Magneto-Optical Trap (MOT) system used to produce Bose-Einstein Condensate (BEC). A primary challenge addressed in this study concerns the geometric limitations of traditional single-pair anti-Helmholtz coil configurations, where the magnetic field peaks occur outside the accessible inter-coil region. To overcome this limitation, we have explored the use of double-pair anti-Helmholtz coil configurations that create well-shaped magnetic field potentials centered at the experimentally accessible $z=0$ location. This investigation encompasses the three sequential processes of atom cooling: cooling in a linear external magnetic field through Doppler cooling, cooling in a well-shaped magnetic field through trapping, and evaporative cooling of atoms to achieve sub-microkelvin temperatures. Through theoretical analysis and numerical simulation, we have determined optimal geometric parameters for the coil configuration and operational parameters including laser detuning, saturation intensity, and initial atom populations for ${}^{87}\text{Rb}$ BEC production. The results indicate that with the optimized configuration, the system can achieve final temperatures of approximately $T_f \sim 60\,\mathrm{nK}$ and produce condensate populations of $\sim 10^5$ atoms with a mean density of $n_0 = 4.9 \times 10^{15}\,\mathrm{m}^{-3}$, providing systematic design guidance for experimental BEC systems.
\end{abstract}

\maketitle

\begingroup
\renewcommand\thefootnote{}
\endgroup

\bigskip

\section{Introduction}

Bose-Einstein Condensates (BECs) were theoretically predicted by Bose and Einstein in 1925. In 1995, the Bose--Einstein condensate was created experimentally for the first time by Eric Cornell and Carl Wieman using rubidium atoms. Later that year, Wolfgang Ketterle from MIT produced a BEC using sodium atoms. In 2001, Cornell, Wieman, and Ketterle shared the Nobel Prize in Physics for the achievement of Bose--Einstein condensation in dilute gases of alkali atoms and for early fundamental studies on the properties of the condensate~\cite{PTB_BEC_lecture_notes_2019}.

The experimental setup for creating BEC is composed of a vacuum chamber, anti-Helmholtz coils, and circularly polarized, frequency-detuned lasers. The coils supply an external magnetic field, and the lasers are shone orthogonal to each other. This setup is a Magneto-Optical Trap (MOT)~\cite{Metcalf1999,Tarbutt2014Magneto-optical}. The magnetic field helps create a trap for the atoms, while the spontaneous absorption and emission of photons cool down the atoms in the trap~\cite{Metcalf1999}. Nowadays, BEC experiments help us further understand the quantum world. Although BEC systems are solely for research purposes, many believe they can help us develop new techniques applicable to the industry~\cite{PTB_BEC_lecture_notes_2019}.

This paper will delve into suggestions regarding the geometric parameters of the MOT~\cite{Pandey2023MOT,Tarbutt2014Magneto-optical} and will use the recommended parameters in simulations of temperature and density of a rubidium (${}^{87}\text{Rb}$) BEC. In this work, there will be three processes of atom cooling investigated: cooling in a linear external magnetic field, cooling in a well-shaped magnetic field, and evaporative cooling of atoms. These processes will be investigated in order to find optimal parameters regarding the setup. Then, the variation of temperature with respect to time will be plotted, as well as the phase space density.

\section{Mathematical Investigation of the BEC-Producing Setup}

\subsection{The Single-Pair MOT Magnetic Field Form and Its Limitations}

The traditional MOT consists of two oppositely polarized coils, meaning they carry currents in opposing directions. These are called anti-Helmholtz coils and have many applications. The anti-Helmholtz coils each carry a current of $i$, have a radius of $R$, have a separation of $2d$ between them, and are wound $\mathcal{N}$ times. They create a magnetic field in the form of Eq.~\eqref{eq:B_field}.

\begin{equation}
B(z) = \frac{\mu_{0} \mathcal{N} i R^{2}}{2} 
\left[
\frac{1}{\left(R^{2} + (z - d)^{2}\right)^{3/2}} 
- 
\frac{1}{\left(R^{2} + (z + d)^{2}\right)^{3/2}}
\right]
\label{eq:B_field}
\end{equation}

Here $z$ is the distance on the axis of the investigated point to the center of the MOT system. Let us denote $I=\mathcal{N}i$ as the total current in the wires. There are two peaks in this field, which can be used as potential energy wells. The condensate can be trapped around the vicinity of the peaks of the magnetic field. These points are found by setting the derivative of $B$ equal to zero, that is $dB/dz=0$. Although these peaks provide sufficient potential wells for trapping neutral atoms, they also yield some problems. When the local peak at the magnetic field of a single-pair anti-Helmholtz coil is used for the trap, it yields a problem: the peak points are outside the region between the coils where the vacuum chamber will be placed. These problems and possible solutions to them will be discussed in the following section.

\subsubsection{Double-Pair Anti-Helmholtz Configuration as Solution}

In order to obtain a setup where the local minima will be inside a reasonable regime, one can consider using two pairs of anti-Helmholtz coils. These coils will share the same axis and be oriented in opposite polarizations. Such that if the first coil carries a current going clockwise, the second one will carry a counterclockwise current. Similarly, the third coil will carry a counterclockwise current while the fourth carries a clockwise one. Therefore, we have two possible configurations regarding the direction of the current flowing in the wires: $(+, -, -, +)$ or $(-, +, +, -)$. This configuration creates a potential well around $z=0$, which is inside the vacuum chamber. 

\begin{figure}[ht]
\centering
\includegraphics[scale=0.27]{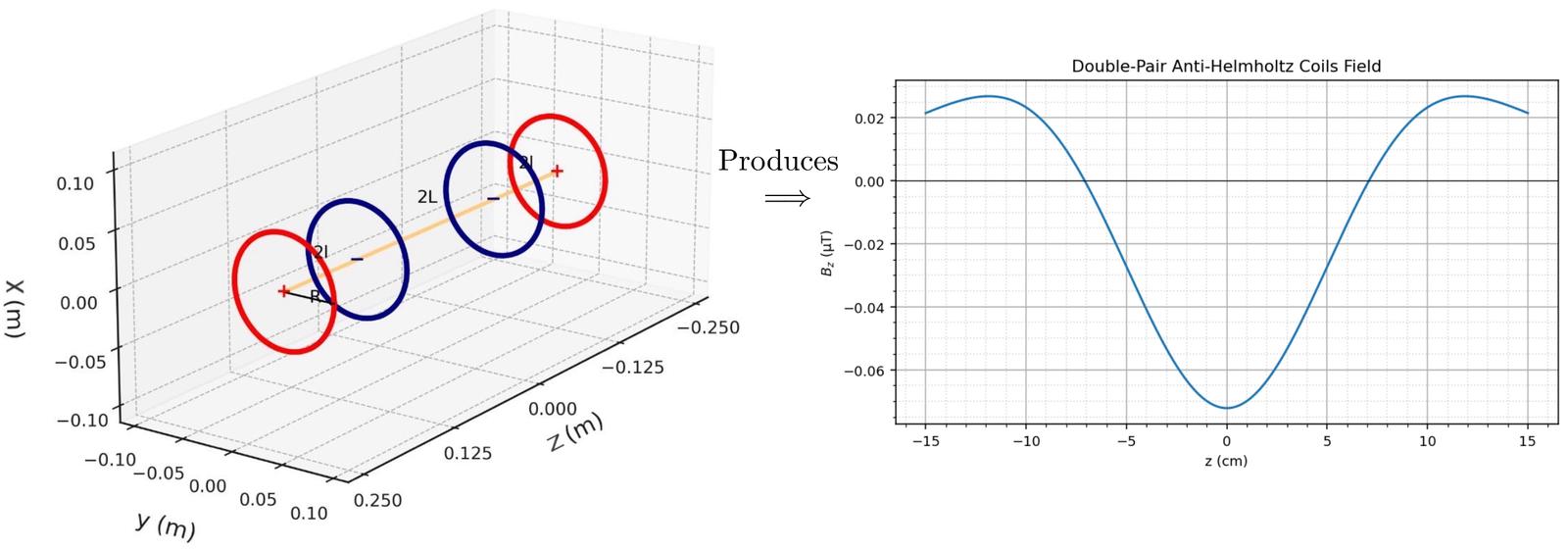}
\caption{Schematic of the Double Anti-Helmholtz Pairs Configuration and the Well-Shaped Magnetic Field.}
\label{fig:mot_system_first}
\end{figure}

We will denote the radii of these four coils as $R$ and the distance between two coils of a pair as $2l$; additionally, the distance between the centers of these pairs will be $2L$. Hence, the coils will be placed at $z=[-l-L, l-L, \allowbreak -l+L, l+L]$. For arbitrary parameters, let us take $I=100\,\mathrm{A}$, $R=l=5\,\mathrm{cm}$ and $L=15\,\mathrm{cm}$. This system and the magnetic field it produces can be visualized with Figure~\ref{fig:mot_system_first} for parameters of $R=10\,\mathrm{cm}$, $l=4\,\mathrm{cm}$, $L=5\,\mathrm{cm}$ and $I=10\,\mathrm{mA}$.

\subsection{Investigation of the Linear Relation Between the External Magnetic Field and the Change in Potential Energy}

In this section the relation between the external magnetic field and the Zeeman splitting of energy levels will be evaluated. The alkali metal atoms being investigated have one valence electron and a stable nucleus. The subatomic particles inside have intrinsic angular momentum called spin. In a simplified case these spins can be modeled to have a linear relation with a magnetic dipole moment. This dipole vector is found by scalarly multiplying a gyromagnetic ratio with the spin vector, so we say $\vec{\mu}=\gamma_G \vec{S}$. In the MOT system these dipoles interact with the magnetic field. The interaction potential energy between dipoles and the field takes form $U=-\hat{\mu} \cdot \vec{B}$. Where the magnetic moment operator is given by Eq.~\eqref{eq:mag_moment}.

\begin{equation}
\hat{\boldsymbol{\mu}} = -\frac{\mu_B}{\hbar} \left( g_L \hat{\mathbf{L}} + g_S \hat{\mathbf{S}} \right) + \frac{\mu_N}{\hbar} g_I \hat{\mathbf{I}}
\label{eq:mag_moment}
\end{equation}

Here $\hat{L}$, $\hat{S}$ and $\hat{I}$ are the operators for the angular momentum, electron spin, and nuclear spin, respectively. The value of $g_I$ is very small, so it is generally omitted. Furthermore, the operators $\hat{J}= \hat{L}+ \hat{S}$ and $\hat{F}=\hat{J}+ \hat{I}$ are used. Hence the energy difference of interaction potential energy between the hyperfine-split levels of an orbital takes the form $\hat{\boldsymbol{\mu}} = - g_F \mu_B \hat{\mathbf{F}}$. For the one-dimensional analysis, the magnetic field along the $z$-axis is in question. Therefore, $\mu_z = - g_F \mu_B m_F$~\cite{PTB_BEC_lecture_notes_2019}.

Here the magnetic quantum number $m_F$ can take integer values in the interval $[-F,-F+1,\ldots,F-1,F]$. Additionally, $g_F$ is given by Eq.~\eqref{eq:gF} when omitting $g_I$, as aforementioned.

\begin{equation}
g_F = g_J \frac{F(F+1) + J(J+1) - I(I+1)}{2F(F+1)}
\label{eq:gF}
\end{equation}

where $F=J+I$ and $J=L+S$ (see Appendix A for further notice). Therefore, this energy difference can be modelled to have a linear relation with the magnetic field, assuming the magnetic field is weak. Hence, let $\Delta E=CB$. Here $C=g_F m_F \mu_B$. This relation will be referenced all throughout the paper as well as the sufficient values of $C$ for the traps.

\subsection[Evaluation of Possible $C$ Values]{Evaluation of the Possible Values For the Linear Proportionality Constant \texorpdfstring{$C=\Delta E/B$}{C=Delta E/B}}

In these sections, the energy differences between Zeeman-split energy levels created by the magnetic field will be investigated. This system will be investigated under Zeeman-split calculations because the external magnetic field is of interest. As this field is fairly weak, the relation between the energy difference and field magnitude is assumed to be linear. This is also known as the linear Zeeman regime. Using the $g$-factors given above, possible $C$ values will be computed (see Appendix B for further calculations).

\textbf{Ground State Electrons in $s$ Orbital:}
The possible values of $C$ for unexcited valence electrons in $s$ orbitals are given by Eq.~\eqref{eq:C_ground}.

\begin{equation}
     C = [-1, -0.5, -0.5, 0, 0, 0.5, 0.5, 1] \mu_B
\label{eq:C_ground}
\end{equation}

Therefore, there are 8 possible values of the change in the potential energy caused by the spin-external magnetic field interaction for the unexcited valence electrons.

\textbf{Excited Electrons in $p$ Orbitals:}
The possible values of $C$ for excited valence electrons in $p$ orbitals are given in Table~\ref{tab:cvalues}.

\begin{table}[ht]
\centering
\caption{Possible Values of $g_F m_F = C/\mu_B$ for Excited Atoms Under Hyperfine Splitting Operator $F$}
\begin{tabular}{ |c|c|c|c| } 
\hline
$F$ & $m_F$ & $g_F$ & $m_F g_F$ \\
\hline
\multirow{1}{*}{$0$} & $0$ & $0$ & $0$ \\
\hline
\multirow{3}{*}{$1$} & $-1$ & $-0.1667$ & $0.1667$ \\
& $0$ & & $0$ \\
& $1$ & & $-0.1667$ \\
\hline
\multirow{5}{*}{$2$} & $-2$ & $0.1667$ & $-0.334$ \\
& $-1$ & & $-0.1667$ \\
& $0$ & & $0$ \\
& $1$ & & $0.1667$ \\
& $2$ & & $0.334$ \\
\hline
\multirow{7}{*}{$3$} & $-3$ & $0.667$ & $-2$ \\
& $-2$ & & $-1.334$ \\
& $-1$ & & $-0.667$ \\
& $0$ & & $0$ \\
& $1$ & & $0.667$ \\
& $2$ & & $1.334$ \\
& $3$ & & $2$ \\
\hline
\end{tabular}
\label{tab:cvalues}
\end{table}

Here, $C=g_F m_F\mu_B$; therefore, there are a total of 24 possible values and 15 distinct values including the $C=0$.

\subsubsection{Finalizing the Discussion on Possible $C$ Values}

There are different suitable values of $C$ for the linear and well-shaped magnetic fields. For the linear magnetic field configuration, $C$ can take both positive and negative values. Furthermore, there are two distinct setups discussed that produce a well-shaped field: single-pair anti-Helmholtz coils and double-pair anti-Helmholtz coils. For the former setup only negative $C$ values will be sufficient for a trapping potential, whereas for the latter case positive values are needed for the $(+, -, -, +)$ configuration and negative values are needed for the $(-, +, +, -)$ configuration. 

Additionally, for any case, it is optimal to choose the maximum or the minimum value, as they supply a better trap potential. However, as the atoms have a random possibility to have any $|F,m_F \rangle$ state, it might be extremely difficult to pinpoint an exact value of $C$. Thus, let us consider the negative average value of $C$ for each orbital. On the ground state $S_{1/2}$, $C_-^{s}=[-1 , -0.5,-0.5]\mu_B$. And for the $P_{1/2}$ and $P_{3/2}$, the values are numerous. The average is given by Eq.~\eqref{eq:C_avg}.

\begin{equation}
C_-^{p}=[-2, -1.334, -0.667, -0.334, -0.1667, -0.1667]\mu_B
\end{equation}

Thus:

\begin{equation}
  \frac{\langle C_-\rangle}{\mu_B}=-\frac{2+1.334+0.667+2 \times 0.334+1+2 \times 0.5}{7}=-\frac{20}{21}
\label{eq:C_avg}
\end{equation}

These values will be referred to as $C$ to decrease the symbolic weight of future calculations.

\subsection{Derivation of the Time-Dependent Differential Equation for the Temperature in a MOT}

At the start of producing a BEC, the system will cool down the atoms by Doppler cooling. This is often done at the center of the MOT, where the magnetic field has a constant gradient. This process is generally for slowing down the atoms before trapping them. After this process is completed, they reach temperatures on the order of microkelvins. Let us find a general equation of motion that is applicable anywhere in the MOT system. Firstly, the force on a particle in a particle and subject to laser light is given by Eq.~\eqref{eq:MOT_force}~\cite{Pandey2023MOT}.

\begin{equation}
    \mathbf{F}_{\text{MOT}} = -\alpha \mathbf{v} - \frac{\alpha g_F m_F\mu_B}{\hbar k} \mathbf{r} \nabla \|\mathbf{B}\|,
    \label{eq:MOT_force}
\end{equation}

where $\alpha=4\hbar k^2 \frac{2s|\delta|/\Gamma}{(1+s+(2\delta/\Gamma)^2)^2}$.

Here $k=\frac{2\pi}{\lambda}$ is the wavenumber of the laser light and $\delta$ is laser detuning~\cite{Tarbutt2014Magneto-optical}. The laser detuning $\delta=\omega_L-\omega_0$ is defined to be the difference between the laser light angular frequency and the angular frequency that corresponds to the energy gap for electron excitation. Let us imagine a single-step energy transition for a single valence electron. In the classical theory of the Bohr atom model, an electron can only receive a specific amount of energy in order to jump this energy gap. However, nowadays we know that this previously predicted energy difference is the middle point of an interval. Let us denote the photons that occupy this middle-point energy to have the specific angular frequency $\omega_0$. This interval is expressed in terms of the angular linewidth $\Gamma$ as $[\omega_0-\Gamma,\omega_0+\Gamma]$. Additionally, the saturation coefficient $s=I/I_{\text{sat}}$ depends on the laser light intensity, where the saturation intensity is given by $I_{\text{sat}}= \frac{\pi h c \Gamma}{3 \lambda^3}$.

Besides this force, atoms also experience the spontaneous force of the random photon absorption and emission process. This force can be formulated by a Dirac delta function at a random time value. Additionally, using the definition of $C$, the equation of motion is an unhomogeneous damped oscillation differential equation in the form of Eq.~\eqref{eq:motion}.

\begin{equation}
 m \ddot{\vec{r}} = -\alpha  \vec{v} - \frac{\alpha C}{\hbar k} \nabla \|  \vec{B} \|  \vec{r} + \vec{F}_{\text{sp}}(t)
    \label{eq:motion}
\end{equation}

which yields

\begin{equation}
\ddot{\vec{r}}+2\gamma \dot{\vec{r}}+\Omega^2 \vec{r}=\vec{a}_{\text{sp}}(t)
\end{equation}

where $2\gamma=\alpha/m$, $\Omega^2=\frac{\alpha C}{m\hbar k} \nabla \| \vec{B} \|$ and $a_{\text{sp}}=F_{\text{sp}}/m$.

In this equation the spontaneous acceleration is modeled as stochastic blows to the velocity of the particle. Specifically, the photons emitted from the atom will have similar wavenumbers as the original laser wavenumber. Therefore, let us say that they will have a momentum of $p=\hbar k_L$, where $k_L$ is the laser wavenumber vector. Thus, a momentum change of $p$ will give the atom a quick blow of velocity $\Delta v= p/m=\frac{\hbar k_L}{m}$. Additionally, these impulses are very quick, so essentially no time is assumed to pass during them. Hence, for stochastic blows $\vec{v}(t_i^+)=\vec{v}(t_i^-)\pm \Delta \vec{v}$. In the next section the general form of the equation of motion will be re-evaluated for specific conditions.

\subsubsection{Evaluation of The Harmonic Term $\Omega$}

As aforementioned, $\Omega^2= \frac{\alpha C}{m\hbar k} \nabla \| \vec{B} \|$. By Eq.~\eqref{eq:B_field}, the derivative of the magnetic field created by single-pair anti-Helmholtz coils is given by Eq.~\eqref{eq:dBdz}.

\begin{equation}
\nabla\|\vec{B}\|=
\frac{dB}{dz} = 
\frac{3 \mu_0 I R^2}{2}
\left[
\frac{z + d}{(R^2 + (z + d)^2)^{5/2}}
-
\frac{z - d}{(R^2 + (z - d)^2)^{5/2}}
\right]
\label{eq:dBdz}
\end{equation}

Here $2d=2(L-l)$ is the separation between the two coils around the center, $z=0$. In order to deal with this long expression, let us expand the derivative through Taylor series around $z=z'$. This is given by Eq.~\eqref{eq:taylor_expansion}.

\begin{multline}
\frac{dB}{dz}=\frac{dB}{dz}\bigg|_{z'}+\frac{d^2B}{dz^2}\bigg|_{z'}(z-z')+\frac{1}{2}\frac{d^3B}{dz^3}\bigg|_{z'}(z-z')^2+\cdots \\
+\frac{1}{n!}\frac{d^{n+1}B}{dz^{n+1}}\bigg|_{z'}(z-z')^n+\cdots
\label{eq:taylor_expansion}
\end{multline}

This $z'$ value will be attributed to special locations in the MOT. For a potential well (that is, the dip from a single-pair anti-Helmholtz or the center of the double-pair anti-Helmholtz coils) $z'$ is the location satisfying $\frac{dB}{dz}\big|_{z'}=0$. Therefore, the first term in the Taylor expansion disappears.

\subsubsection{Evaluation of the Parameters for Stochastic Photon Recoil Impulses and Lasers}

In this part the average frequency of the aforementioned random impulses will be investigated. In order to find the rate at which the stochastic blows of velocity happen, let us investigate the scattering rate of photons. That is, how many photons are absorbed or emitted in a second, and what their impact is on the velocity of atoms. The total scattering rate of photons from both light beams in opposite directions is given by Eq.~\eqref{eq:Rsc}.

\begin{equation}
R_{\mathrm{sc}} = \frac{1}{2}
\frac{\Gamma s}{1 + s + \left( {2\delta}/{\Gamma} \right)^2},
\label{eq:Rsc}
\end{equation}

This defines how many, essentially random, impulses will be given to the subject atoms. Now, let us investigate the impulse recoils the atoms are subject to at a rate of $R_{\text{sc}}$.

The process of absorption and emission of photons is essentially random; therefore, $\langle\Delta\vec{p}\rangle=0$. The vector property of these random impulses makes them average to zero when calculated over a long time duration. Thus, let us investigate the root-mean-square of their amplitude using the Doppler optical molasses cooling theory. For now, we will disregard the magnetic field, as it does not contribute to the solution. The photon recoils are assumed to happen at a Poisson rate; that is, the probability of a photon recoil taking place in a time interval $dt$ is $P=R_{\text{sc}}dt$. Hence the probability of a recoil not taking place is $Q=1-P=1-R_{\text{sc}}dt$. Therefore, there is a $Q$ probability that $\vec{p}\rightarrow\vec{p}-\alpha\vec{v} dt/m$ and a $P$ probability that $\vec{p}\rightarrow\vec{p}-\alpha\vec{p} dt/m\pm\Delta\vec{p}$. Thus, $p^2\rightarrow p^2-2\alpha p^2 dt/m+\alpha^2p^2dt^2/m^2$ or $p^2\rightarrow p^2-2\alpha p^2 dt/m+\alpha^2p^2dt^2/m^2+p(1-\alpha dt/m)\Delta p+(\Delta p)^2$ will take place with respective probabilities of $Q$ and $P$. Let us omit the terms in the second order of $dt$ and take the average. Let us recall that the vector $\vec{p}$ changes its direction and its mean value is zero. So, $\langle p^2\rangle\rightarrow\langle p^2\rangle-2\alpha\langle p^2\rangle dt/m$ or $\langle p^2\rangle\rightarrow\langle p^2\rangle-2\alpha\langle p^2\rangle dt/m+\langle(\Delta p)^2\rangle$. Combining the two possibilities: 

\begin{align}
\langle p^2\rangle &\rightarrow(\langle p^2\rangle-2\alpha\langle p^2\rangle dt/m)(1-R_{\text{sc}}dt) \\
&\quad +(\langle p^2\rangle-2\alpha\langle p^2\rangle dt/m+\langle(\Delta p)^2\rangle)R_{\text{sc}}dt \\
&= \langle p^2\rangle-2\alpha\langle p^2\rangle dt/m+\langle(\Delta p)^2\rangle R_{\text{sc}}dt
\end{align}

Therefore, $d(\langle p^2\rangle)=-2\alpha\langle p^2\rangle dt/m+\langle(\Delta p)^2\rangle R_{\text{sc}} dt$, which gives

\begin{equation}
\frac{d}{dt}\langle p^2\rangle=-\frac{2\alpha}{m}\langle p^2\rangle+\langle(\Delta p)^2\rangle R_{\text{sc}}
\end{equation}

Substituting this into $T=\frac{\langle p^2\rangle}{mk_B}$ the equation for the temperature is found as 

\begin{equation}
\frac{dT}{dt}=-\frac{2\alpha}{m} T+\frac{ R_{\text{sc}}}{k_B}\langle(\Delta p)^2\rangle
\end{equation}

In the steady state the temperature, on average, stays constant; therefore $\frac{dT}{dt}\big|_{t\rightarrow\infty}=0$. Hence, $T_\infty$ is given by Eq.~\eqref{eq:T_infinity}~\cite{Metcalf1999}.

\begin{equation}
    T_{\infty}=\frac{\langle(\Delta p)^2\rangle R_{\text{sc}}}{2\alpha k_B}=\frac{\hbar \Gamma}{2k_B}\frac{1+s+(2\delta/\Gamma)^2}{4|\delta|/\Gamma}
\label{eq:T_infinity}
\end{equation}

Therefore, the final temperature is in linear relation with the saturation parameter $s$. Thus, let us choose $s=0.01$ so the final steady state temperature $T_\infty\propto(1.01+4\beta^2)/4\beta\approx(1+4\beta^2)/4\beta$. Now, let us find the optimal value of $\beta$ by setting $\frac{\partial T_\infty}{\partial\beta}=0$. This gives $32\beta^2-4(1+4\beta^2)=0\Rightarrow\beta=\pm0.5$. As aforementioned, let us choose the lasers to be red-detuned; therefore $\beta=0.5$.

\subsubsection{Temperature Evolution in the First Setup: Linear Magnetic Field Doppler Cooling}

The linear magnetic field is generally used for the initial cooling of the atoms. With the classical laser cooling setup, atoms can reach temperatures on the order of microkelvins in the vicinity of the center of the single-pair MOT system. Now, let us find $\Omega$ in this configuration. Only the first derivative of the magnetic field will be used since $z\approx0$, a region where the field is quite linear. Therefore, the $\Omega_1$ of the system is given in Eq.~\eqref{eq:omega1}.

\begin{equation}
    \frac{dB}{dz}\bigg|_{z=0}=\pm\frac{3\mu_0IR^2}{(R^2+l^2)^{5/2}}\Longrightarrow \Omega_1^2= \frac{\alpha C}{m\hbar k}  \frac{3\mu_0IR^2}{(R^2+l^2)^{5/2}}
\label{eq:omega1}
\end{equation}

Having found the equation of motion for the atoms, the temperature evaluation with respect to time will be modeled. The temperature will be calculated according to one-dimensional Boltzmann statistics; that is $T=\frac{m\langle v^2\rangle}{k_B}$. For now, a small value for the initial temperature will be used in order to make it easier to see the cooling-down process. The laser is often red-tuned; therefore, the detuning $\delta$ is negative. Let us define $\beta=-\delta/\Gamma$. The equation of motion will be run for $N=2000$ atoms (this is very small compared to the needed number of $\sim10^6$; however, this does not change the steady-state temperature; a large atom number only makes the graph run smoother with a smaller error). The average of the squared speed will be evaluated as an arithmetic mean, that is $\langle v^2\rangle=\frac{1}{N}\sum_{i=1}^N v_i^2$. After this calculation the variance in temperature is $\sigma^2=\frac{1}{N}\sum_{i=1}^N (T_i-\langle T\rangle)^2=\frac{1}{N}\frac{m^2}{k_B^2}\sum_{i=1}^N (v_i^2-\langle v^2\rangle)^2$ therefore the standard error is $S=\sigma/\sqrt{N}$. The mass of ${}^{87}\mathrm{Rb}$ atoms is $m=1.443\times10^{-25}\,\mathrm{kg}$, and the angular linewidth of these atoms are $\Gamma=2\pi\times 6.07\,\mathrm{MHz}$. The laser wavelength $\lambda=\frac{2\pi}{k}=780.24\,\mathrm{nm}$ is chosen to suffice for the $\text{D2:}\,5S_{1/2}\rightarrow5P_{3/2}$ transition line of ${}^{87}\text{Rb}$.

\begin{figure}[ht]
    \centering
    \includegraphics[scale=0.6]{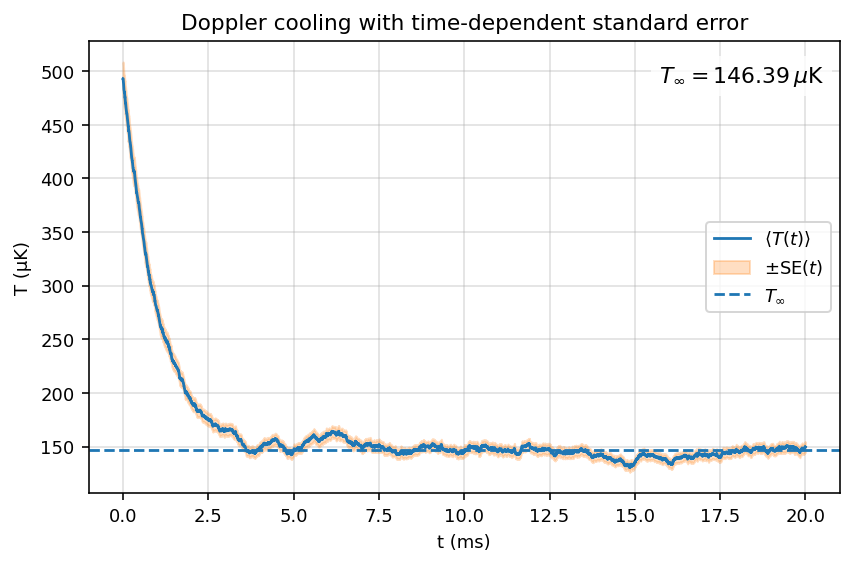}
    \caption{Plot of Temperature vs. Time With Shaded Area for the Interval Inside One SE From the Average Value.}
    \label{fig:temperature_linear}
\end{figure}

Figure~\ref{fig:temperature_linear} depicts the temperature change with respect to time starting from a mean temperature of $T_0=0.5\,\mathrm{mK}$. The blue dashed line shows the steady-state value of the temperature; in this case the predicted Doppler temperature is $T_\infty=146.4\,\mathrm{\mu K}$.

\subsubsection{Second Setup: Well-Shaped Magnetic Field Trapping of Neutral Atoms}

Now, the second stage of the setup will be investigated: the temperature variation in the magnetic trap. Most notably, the difference between the first and second setups is the derivative of the magnetic field. As a well-shaped magnetic field will be used here, the first derivative of the magnetic field around $z'$ will be zero, regardless of whether $\frac{dB}{dz}\big|_{z=z'}=0$ for the single anti-Helmholtz coil pair or $z'=0$ for the double anti-Helmholtz coil pair is used. Therefore, the second term in the Taylor expansion given in Eq.~\eqref{eq:taylor_expansion} is used. Hence, Eq.~\eqref{eq:omega2_def} is obtained for the so-called ``angular frequency'' of the second setup.

\begin{equation}
\Omega^2=\frac{\alpha C}{m\hbar k}  \frac{d^2B}{dz^2}\bigg|_{z'}(z-z') + \mathcal{O}[(z-z')^2]
\label{eq:omega2_def}
\end{equation}

Substituting into the equation of motion gives Eq.~\eqref{eq:motion_well}.

\begin{equation}
\ddot{\vec{r}}+\frac{\alpha}{m} \dot{\vec{r}}+\frac{\alpha C}{m\hbar k}  \frac{d^2B}{dz^2}\bigg|_{z'}(z-z') \vec{r}+ \mathcal{O}[(z-z')^2] \vec{r}=\vec{a}_{\text{sp}}(t)
\label{eq:motion_well}
\end{equation}

Here $\vec{r}=(z-z')\hat{z}$. Let us omit the terms in higher orders of $r$. Recalling the previously defined parameters, Eq.~\eqref{eq:motion_simple} is obtained.

\begin{equation}
\ddot{r}+2\gamma \dot{r}+\Omega_2^2 r =  a_{\text{sp}}(t)
\label{eq:motion_simple}
\end{equation}

From now on, the single-pair anti-Helmholtz coil will not be of interest regarding its difficulty in obtaining a well-shaped magnetic field. Specifically, the tip of a dip at the function created by the single-pair system is outside the region confined between the two coils, that is $z'\notin[-(L-l),L-l]=[-1,1]\,\mathrm{cm}$. This creates an experimental difficulty, as the vacuum chamber in which the rubidium atoms are fired will also have to contain one of the coils in it. This is not an ideal configuration; hence, it will be neglected for the time being.

Furthermore, the second derivative of the magnetic field created by double-pair anti-Helmholtz coils in a polarization configuration of $(+, -, -, +)$ around $z'=0$ is given by Eq.~\eqref{eq:B_double_prime}.

\begin{equation}
B''(0)
= 3\mu_0IR^2 \left[
\frac{5(L+l)^2 - R^2}{\left(R^2 + (L+l)^2\right)^{7/2}}
- \frac{5(L-l)^2 - R^2}{\left(R^2 + (L-l)^2\right)^{7/2}}
\right]
\label{eq:B_double_prime}
\end{equation}

Let us define $A=L/R$, $a=l/R$, and a function $f(a,A)$ satisfying Eq.~\eqref{eq:f_function}.

\begin{equation}
B''(0)
= \frac{3\mu_0I}{R^3} \left[
\frac{5(A+a)^2 - 1}{\left(1 + (A+a)^2\right)^{7/2}}
- \frac{5(A-a)^2 - 1}{\left(1 + (A-a)^2\right)^{7/2}}
\right]=\frac{3\mu_0I}{R^3}f(a,A)
\label{eq:f_function}
\end{equation}

which yields

\begin{equation}
\Omega_2^2=\frac{\alpha C}{m\hbar k}\frac{3\mu_0I}{R^3}f(a,A)
\end{equation}

See Appendix~C for the investigation of $f(a,A)$ in order to maximize it and consequently find the optimal configuration of the double-pair anti-Helmholtz coils. The MOT system has $R=10\,\mathrm{cm}$, $L=5\,\mathrm{cm}$ and $l=4\,\mathrm{cm}$. Therefore $f(0.4,0.5)=1.3$.

\begin{figure}[ht]
    \centering
    \includegraphics[scale=0.6]{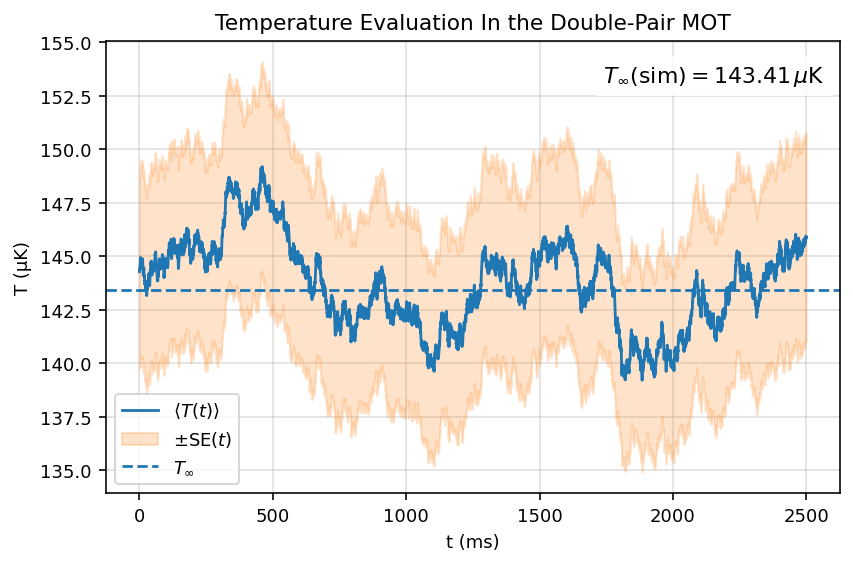}
    \caption{Plot of Temperature vs. Time Averaged Over an Ensemble of Atoms in a Double Pair Coil MOT System}
    \label{fig:temperature_well}
\end{figure}

The most notable difference between the equations of motion in the two systems is the oscillating terms. Now, let us proceed with the same graphing technique we have done previously. This time the parameters of the MOT must also be defined; these are the total current in the coils, $I$, the radius of the coils, $R$, the separation within each pair, $2l$, the separation between the centers of the pairs, $2L$. However, this time the goal is not cooling down the atoms but rather creating a trap which will keep the steady-state temperature somewhat stable and let the atoms evaporate freely off the condensate. Therefore, this time the saturation parameter $s$ is chosen smaller. Figure~\ref{fig:temperature_well} is obtained for the temperature versus time graph to see how the steady-state temperature holds.

Despite some fluctuations on the order of $\sim5\,\mathrm{\mu K}$, the temperature stays pretty stable. However, this system does start to heat up the condensate after three seconds for $s=10^{-5}$. This duration where the temperature changes with $s$ values, lower $s$ values corresponding to larger stable durations. Therefore, after the second coils are operated and one of the central coils is changed in polarization, the laser intensity needs to be lowered to zero, preferably within three seconds. 

From now on the system will act as a quadratic trap with potential in the form $V=CB=CB(0)+\frac{1}{2}CB''(0)z^2$. This traps neutral atoms and prevents Majorana spin flip losses with its constant term. However, this does not trap atoms with $C=0$; therefore, a quarter (6 out of 24 values of $C$ are zero) of the atoms will be shot out. The trapping force is $\vec{F}=-\frac{\partial V}{\partial z}\hat{z}=-CB''(0)\vec{z}$. Thus the equation of motion becomes $m\ddot{\vec{z}}=-CB''(0)\vec{z}\Longrightarrow\ddot{r}+\frac{CB''(0)}{m}r=0$. This is a classic quadratic trap that does not heat nor cool the trapped atoms; however, some atoms are too energetic to be trapped and can escape the trap since it is not a perfect parabola. This is called evaporative cooling and will be investigated in the next section.

\subsection{Reaching the Final Temperature With Evaporative Cooling}

Evaporative cooling is a random process at its core; however, we can still model it using statistical models. Monte-Carlo is one of the most widely used simulation models for high-energy escaping atoms. In our system the potential well depth is $U_0=CB(z')$. The evaporating atoms must have sufficient energy to escape this well. Let us introduce the energy ratio $\eta =\frac{k_BT}{U}$. We can assume that the escaping atoms must have an energy ratio $\eta>\eta_0$ where $\eta_0=\frac{k_BT}{CB(z')}$ is the threshold parameter. In order to find $\eta_0$ let us investigate the maximums and minimums of the magnetic field created by the double-pair anti-Helmholtz coils. Let $\Delta B=\text{max}(B)-B(z=0)$, therefore, the constraint on temperature is given by $T<\frac{C\Delta B}{k_B}$. Now, let us find an appropriate value for $\Delta B$. For ${}^{87}\text{Rb}$ atoms the threshold temperature for BEC formation is around $T\sim150\,\mathrm{nK}$. This corresponds to $\Delta B\sim240\,\mathrm{nT}$; therefore, let us aim for $\Delta B\sim100\,\mathrm{nT}$ which corresponds to $T_f\sim64\,\mathrm{nK}$.

The cooling down process works by shooting out atoms with the highest velocities. Now, let us calculate the fraction of the atoms that stay in the condensate by using Maxwellian probability distribution. Let there be $N$ atoms at the start and $N'$ atoms after the other ones are kicked out. According to Maxwellian distribution, the probability of an atom having a velocity in the interval $[v,v+dv]$ is given by $q=g(v)dv$ where $g(v)=\sqrt{\frac{\kappa}{\pi}}e^{-\kappa v^2}$ with $\kappa=m/2k_BT$. In order to find $N'/N$ let us calculate the probability atoms have a velocity in the interval $[-v_f,v_f]$ where $v_f^2=k_BT_f/m$. Therefore 

\begin{equation}
N'/N=q=\int_{-v_f}^{v_f}g(v)dv=2\sqrt{\frac{\kappa}{\pi}}\int_0^{v_f}e^{-\kappa v^2}dv
\end{equation}

where $\kappa=m/2k_BT_\infty$. This integral is evaluated with the error function; that is, $\text{erf}(x)=\frac{2}{\sqrt{\pi}}\int_0^xe^{-t^2}dt$, which gives 

\begin{equation}
\int_0^{v_f}e^{-\kappa v^2}dv=\frac{1}{2}\sqrt{\frac{\pi}{\kappa}}\text{erf}(\sqrt{\kappa}v_f)
\end{equation}

Thus, 

\begin{equation}
q=\text{erf}(\sqrt{\kappa}v_f)=\text{erf}\left(\sqrt{\frac{m}{2k_BT_\infty}\frac{k_BT_f}{m}}\right)=\text{erf}(\sqrt{T_f/2T_\infty})
\end{equation}

From here the fraction is found to be 

\begin{equation}
q=\text{erf}\left(\sqrt{64\,\mathrm{nK}/2 \times 146\,\mathrm{\mu K}}\right)=\text{erf}(0.0148)=0.0167=1.67\%
\end{equation}

A typical BEC consists of $N'\sim 10^5$ atoms; therefore, the number of atoms that make it to the second setup should be around $N\sim10^7$ (accounting for the $C=0$ losses). Additionally, when switching from the single-pair to the double-pair system half of the atoms will be lost as the double-pair can only support $C>0$ for the $(+, -, -, +)$ polarization and $C<0$ for $(-, +, +, -)$. Either way, there is a loss. Thus the initial number of atoms fired into the MOT must be around $N_0\sim5 \times 10^7$.

\subsection[Derivation of Density Profile]{Derivation of Density Profile Using Thomas--Fermi Approximation}

In order to find a Phase Space Density (PSD) the Thomas--Fermi approximation is used. This approximation neglects the kinetic energy of the atoms (a very small $\eta$ parameter) in the Gross--Pitaevskii Equation (GPE) to find a density profile. The one-dimensional GPE is given by Eq.~\eqref{eq:gpe}.

\begin{equation}
  \left( -\frac{\hbar^2}{2m} \frac{\partial^2}{\partial r^2} + V(\mathbf{r}) + \frac{2 \hbar \omega_\perp a_s}{1-\zeta(1/2)\frac{a_s}{a_\perp}} |\psi(\mathbf{r})|^2 \right) \psi(\mathbf{r}) = \mu \psi(\mathbf{r})
  \label{eq:gpe}
\end{equation}

Here $\mu$ is the chemical potential of the condensate and $a_s$ is the scattering length. The $\omega_\perp$ and $a_\perp$ are the angular frequency of oscillations in the $x$--$y$ plane and the estimated size of the condensate in the $x$--$y$ plane, respectively. For ${}^{87}\mathrm{Rb}$ experimental results yield $a_s^{{}^{87}\mathrm{Rb}}=100a_0$ where $a_0=5.29 \times 10^{-11}\,\mathrm{m}$ is the Bohr radius. The kinetic energy term in the GPE comes from the first term with the spatial derivative. Let us ignore this term, as the atoms have reached incredibly small velocities, and denote 

\begin{equation}
\varrho=\frac{2 \hbar \omega_\perp a_s}{1-\zeta(1/2)\frac{a_s}{a_\perp}}\approx 100\hbar a_0\sqrt{\frac{2|CB''(0)|}{m}}
\end{equation}

(see Appendix~E for derivation)~\cite{Olshanii1998}. Hence, Eq.~\eqref{eq:density} is obtained.

\begin{equation}
n\approx \frac{\mu -V}{\varrho}=\frac{\mu-CB}{\varrho}
\label{eq:density}
\end{equation}

Now, let us say there are $N$ particles in the condensate, and also confine it in the interval $[z'-z_0, z'+z_0]$ with $z_0\ll z'$. As the previously defined density is one-dimensional, the integral is taken over $z$ only.

\begin{equation}
    \int_{z'-z_0}^{z'+z_0} n\,dz= \int_{z'-z_0}^{z'+z_0}  \frac{\mu -CB}{\varrho}dz=N' \Longrightarrow 2\mu z_0-C\int_{z'-z_0}^{z'+z_0} B\,dz=N'\varrho
\end{equation}

The integral of the magnetic field can be estimated under the condition $z_0\ll z'$ as in Eq.~\eqref{eq:B_integral}.

\begin{equation}
\int_{z'-z_0}^{z'+z_0} B(z)\,dz
= 2 z_0 B(z') + \frac{1}{3} z_0^3 B''(z') + \mathcal{O}(z_0^5)
\label{eq:B_integral}
\end{equation}

Therefore, Eq.~\eqref{eq:mu_eq} is obtained.

\begin{equation}
\begin{aligned}
2\mu z_0&=N'\varrho+C \left[2 z_0B(z') + \frac{1}{3} z_0^3 B''(z') + \mathcal{O}(z_0^5)\right] \\
 \mu&=\frac{N'\varrho}{2z_0}+C \left[B(z') +\frac{1}{6} z_0^2 B''(z') + \mathcal{O}(z_0^4)\right] 
\end{aligned}
\label{eq:mu_eq}
\end{equation}

When the atom structure is approaching the BEC configuration, the chemical potential gets minimized. Thus,  $\frac{d\mu}{dz_0}=0$. This determines the gas confinement regime's size, $z_0$. Omitting the higher terms on the order of four or higher in Eq.~\eqref{eq:mu_eq} gives Eq.~\eqref{eq:z0_optimal}.

\begin{equation}
         \frac{d\mu}{dz_0}
= -\frac{N'\varrho}{2 z_0^2} + \frac{1}{3} C z_0 B''(z')=0 \Longrightarrow z_0 = \left(\frac{3N'\varrho}{2C B''(z')}\right)^{1/3}
\label{eq:z0_optimal}
\end{equation}

Substituting this into the chemical potential function, Eq.~\eqref{eq:mu_final} is obtained.

\begin{equation}
    \mu(z_0) = [CB''(z')]^{1/3}\left(\frac{3N'\varrho}{2}\right)^{2/3} + C B(z')
\label{eq:mu_final}
\end{equation}

Lastly, plugging Eq.~\eqref{eq:mu_final} into Eq.~\eqref{eq:density}.

\begin{equation}
    n(z) = \frac{1}{\varrho} \left[
[CB''(z')]^{1/3} \left(\frac{3 N'\varrho}{2}\right)^{2/3} 
+  C B(z')-CB(z) \right]
\label{eq:n_final}
\end{equation}

where $z'=0$ for the double-pair anti-Helmholtz coils configuration.

\begin{figure}[ht]
    \centering
    \includegraphics[scale=0.6]{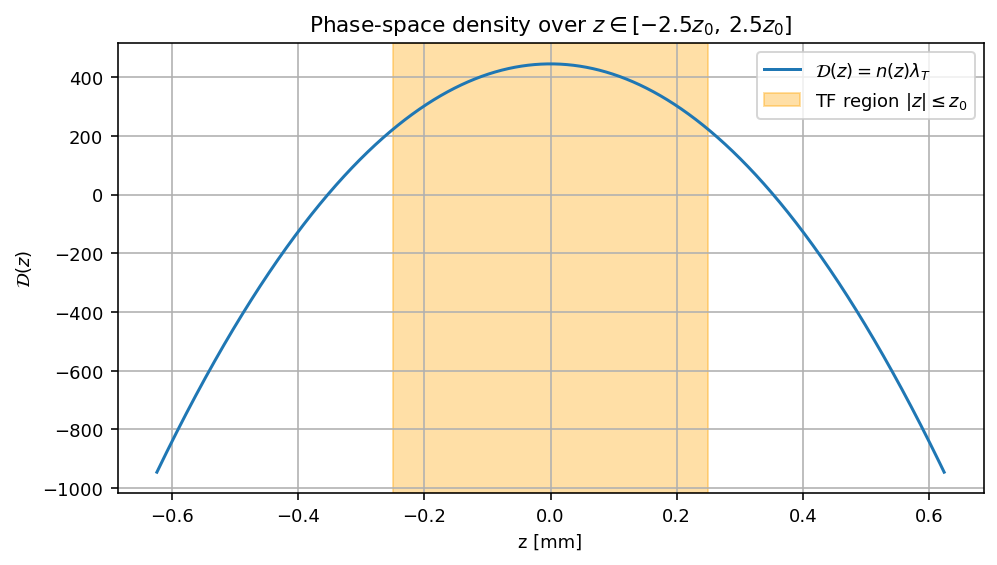}
    \caption{The Phase-Space Density Profile $\mathcal{D}(z)$ vs. $z$ Coordinate with Shaded Region of $[-z_0,z_0]$}
    \label{fig:density_psd}
\end{figure}

In order to obtain a dimensionless parameter, the Phase-Space Density (PSD) is used which is given by $\mathcal{D}(z)=n(z)\lambda_T$ where $\lambda_T = \sqrt{\frac{2\pi\hbar^2}{mk_B T}}$ is the thermal wavelength. Therefore, Figure~\ref{fig:density_psd} is obtained.

\section{Conclusion}

Throughout this paper, suggestions were made regarding the geometric parameters of the coil configuration and other parameters such as initial atom number, current through the coils, and laser intensity. It was decided that the experimental system will be set up as follows in order to obtain the middle point of wins and losses, optimizing the whole procedure.

First, two coils of radius $R=10\,\mathrm{cm}$ are placed $2d=2\,\mathrm{cm}$ away from each other with opposite polarizations; that is, the current through one is clockwise while the other is counterclockwise. Then three mirrors and three lasers with detuning $\delta=-\Gamma/2$ and saturation $s=0.01\Rightarrow I=\frac{\pi hc\Gamma}{300\lambda^3}=0.167\,\mathrm{Wm^{-2}}$ will be placed to supply an orthogonal set of light beams in all six directions. A vacuum chamber of maximum volume of $14\,\mathrm{cm}\times 14\,\mathrm{cm}\times\text{some length}$ will be placed inside the two central coils. And lastly, the somewhat previously cooled rubidium atoms will be fired into the MOT. There needs to be around $N_0\sim5 \times 10^7$ initial atoms fired into the vacuum chamber. When the atoms are in the chamber, the devices which fire ${}^{87}\text{Rb}$ are removed to make space for the second pair of coils.

Secondly, after the Doppler cooling in the single-pair anti-Helmholtz coil system is completed, which takes around $10\,\mathrm{ms}$, the second pair of coils of radius $R=10\,\mathrm{cm}$ are placed $2l=8\,\mathrm{cm}$ away from the closest central coil. These coils need to have the same polarization. Then the central coil that has the same polarization as the secondary coils must be changed to supply a current in the opposite direction. Therefore, a double-pair anti-Helmholtz coil system is obtained. Now, the intensity of laser light needs to be dimmed, preferably within three seconds. If the condensate gets hotter during this stage the BEC threshold temperature can still be reached; however, more atoms will be fired out in the evaporative cooling stage.

Finally, when the lasers are turned off, the atoms are left to evaporate off the quadratic potential well in Figure~\ref{fig:mot_system_first}, reaching a temperature around $T_f\sim60\,\mathrm{nK}$. In the end a condensate cloud of approximate volume of an ellipsoid with dimensions $\rho=a_\perp$ and $z=z_0$. Therefore, the approximate volume $V\approx\frac{4\pi}{3}a_\perp^2z_0=0.02046\,\mathrm{mm^3}$ (see Appendix~D for $a_\perp$) contains $N\sim10^5$ atoms, which corresponds to a mean density of $n_0=4.9 \times 10^{15}\,\mathrm{m^{-3}}$. Although this seems dilute for a classic BEC experiment with mean densities on the order of $10^{18}\,\mathrm{m^{-3}}$, this can be increased simply by increasing the initial atom number.

\section*{Appendix}

\appendix
\section{\texorpdfstring{Landé $g$-factors of Sub-atomic Parameter Operators}{Lande g-factors of Sub-atomic Parameter Operators}}

The equations (5) and (6) can be found with the given subatomic parameters and their $g$-factors. These parameters are the angular momentum operator $J=L+S$ where $L$ is the electronic angular momentum and $S$ is the electronic spin; furthermore, $I$ is the nuclear spin. For Rubidium atoms $I=3/2$. The Landé factors for $J$ and $F$ are given as:

\begin{equation}
g_J = g_L \frac{J(J+1) + L(L+1) - S(S+1)}{2J(J+1)}
    + g_S \frac{J(J+1) - L(L+1) + S(S+1)}{2J(J+1)}
\end{equation}

and

\begin{equation}
g_F = g_J \frac{F(F+1) + J(J+1) - I(I+1)}{2F(F+1)}
     + g_I \frac{\mu_N}{\mu_B} \frac{F(F+1) + I(I+1) - J(J+1)}{2F(F+1)}
\end{equation}

Here the second term in $g_F$ will be ignored, as $\mu_N/\mu_B$ is proportional to the ratio of $m_e/m_p\approx1/2000$. Therefore, the following is obtained:

\begin{equation}
g_F = g_J \frac{F(F+1) + J(J+1) - I(I+1)}{2F(F+1)} 
\end{equation}

\section{\texorpdfstring{Calculations of the Possible $C$ Values}{Calculations of the Possible C Values}}

\subsection{\texorpdfstring{Ground State Valence Electrons in $s$ Orbitals}{Ground State Valence Electrons in s Orbitals}}

For orbitals $nS_{1/2}$, the angular momentum quantum number $l=0$ so, $J=1/2$. Thus, $g_J=g_S \approx 2$~\cite{Mohr2016CODATA}. For alkali metals like sodium and rubidium the nuclear spin operator $I=3/2$. Therefore, $F$ has two possible values as $F=[1, 2]$. Hence, the following is obtained from the formula for $g_F$:

\begin{align}
g_F(F=1) &= 2 \frac{2 + \frac{3}{4} - \frac{15}{4}}{4} = -0.5 \\
g_F(F=2) &= 2 \frac{6 + \frac{3}{4} - \frac{15}{4}}{12} = 0.5
\end{align}

Therefore the possible values for the magnetic quantum number are:

\begin{align}
m_F(F=1) &= [-1, 0, 1] \\
m_F(F=2) &= [-2, -1, 0, 1, 2]
\end{align}

Thus, the possible values of $g_F m_F$ are:

\begin{align}
[g_F m_F](F=1) &= [0.5, 0, -0.5] \\
[g_F m_F](F=2) &= [-1, -0.5, 0, 0.5, 1]
\end{align}

which yields

\begin{equation}
C = [-1, -0.5, -0.5, 0, 0, 0.5, 0.5, 1] \, \mu_B
\end{equation}

\subsection{\texorpdfstring{Excited Valence Electrons in $p$ Orbitals}{Excited Valence Electrons in p Orbitals}}

When the valence electron in alkali metals gets excited, it jumps into an orbital where the angular momentum quantum number is no longer zero but instead $l=1$. This new complexity adds new energy levels into the orbitals. The previously discussed main orbital energies undergo more Zeeman splitting regarding the inner workings of the atom in question, such as spin vector configuration. The transition from an $s$ orbital to a $p$ orbital has 2 possible transitions: one with a resultant $J=1/2$ and the other with a resultant $J=3/2$. These two different possibilities are represented by the D lines. The D lines for Sodium (${}^{23}\mathrm{Na}$) and Rubidium (${}^{87}\mathrm{Rb}$) are given in Table~\ref{tab:dlines}.

\begin{table}[ht]
\centering
\caption{D-line Transitions and the Wavelength Required for Them in Sodium and Rubidium}
\begin{tabular}{||c c c c||} 
 \hline
Element & D-line & Transition & $\lambda$ (nm) \\
 \hline\hline
 ${}^{23}\mathrm{Na}$ & $D_1$ & $3S_{1/2} \rightarrow 3P_{1/2}$ & $589.6$ \\
 \hline
 ${}^{23}\mathrm{Na}$ & $D_2$ & $3S_{1/2} \rightarrow 3P_{3/2}$ & $589.0$ \\
 \hline
 ${}^{87}\mathrm{Rb}$ & $D_1$ & $5S_{1/2} \rightarrow 5P_{1/2}$ & $794.8$ \\
 \hline
 ${}^{87}\mathrm{Rb}$ & $D_2$ & $5S_{1/2} \rightarrow 5P_{3/2}$ & $780.2$ \\
 \hline
\end{tabular}
\label{tab:dlines}
\end{table}

Now, in order to find the possible values for $C$ the values of $g_J$, and consequently, the values of $g_F$ for differing $J$ and $F$ values must be calculated. In this case $L=1$, $S=1/2$ and $J=L\pm S$, additionally, $F=J\pm I$. The indexes under the orbitals suggest the $J$ number. While $J$ can take the values $[1/2,3/2]$, the $I$ value is set at $3/2$. This gives the possible outcomes $F=[0,1,2,3]$. When $g_L=1$ and $g_S\approx 2$~\cite{Mohr2016CODATA} are used:

\begin{align}
g_J(J=1/2)&=\frac{1}{6} \\
g_J(J=3/2)&=\frac{2}{3}
\end{align}

Thus, the possible values of $C$ for excited valence electrons in $p$ orbitals are obtained as shown in Table~\ref{tab:cvalues}.

\section{Discussion of the Coil Configuration Geometric Parameters}

\begin{figure}[ht]
    \centering
\includegraphics[scale=0.65]{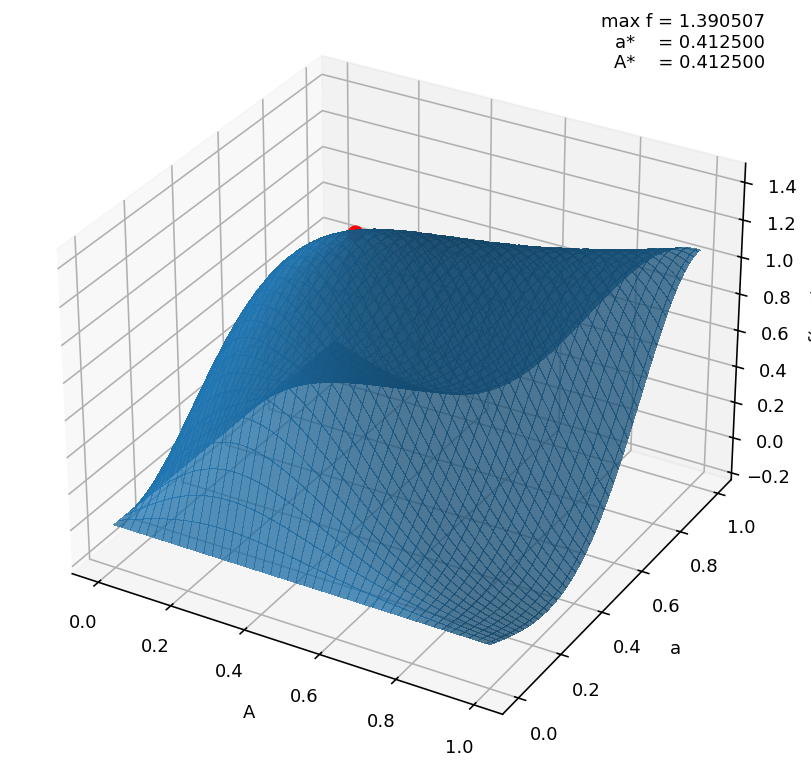}
    \caption{3D Plot of $f(a,A)$ With Its Global Maximum Point}
    \label{fig:f_optimization}
\end{figure}

Figure~\ref{fig:f_optimization} shows the 3-D graph of $f(a,A)$ and its maximum value along with the $a^*$ and $A^*$ values which satisfy $f(a^*,A^*)=\text{max}(f)$.

The graph shows $a^*=A^*$, but this is not physically logical to implement into the system. Thus, let us define $\varepsilon=\frac{L-l}{R}=A-a$. The graph suggests $\varepsilon\ll1$, however, let us choose $\varepsilon=0.1$; therefore, $L=\frac{R}{10}+l$. Now, let us choose $a=0.4$ and consequently $A=0.5$ in order to stay around the regime of the point $(a^*,A^*)$. With these parameters, $f(0.4,0.5)=1.3$ which is quite close to the maximum value of $1.391$. 

Now, recall $\Delta B\sim100\,\mathrm{nT}$. The suitable $R$ and $I$ values need to suffice for a field that is steep enough for a small $z_0$ hence a more compact condensate, but also a field with a small enough depth $\Delta B$ so the final temperature can drop below the BEC threshold. The second derivative of the field has the relation $B''(0)\propto I/R^3$ and the field depth $\Delta B\propto I/R$. Therefore, the radius $R$ has to be as small as possible but also has to fit the vacuum chamber inside of it. Let us choose $R=10\,\mathrm{cm}$, this way there is room to fit a vacuum chamber of size $14\,\mathrm{cm}\times 14\,\mathrm{cm}\times\text{some length}$. 

Now, finally, let us choose an $I$ value. By graphing the field depth for various $R$ values the field can be empirically calculated as $\Delta B(\mathrm{nT})=990.5 I(\mathrm{A})/R(\mathrm{m})$. Therefore $I\sim10\,\mathrm{mA}$. This gives $B''(0)=\frac{3\mu_0I}{R^3}f(a,A)\sim50\,\mathrm{\mu T/m^{2}}$.

\section{\texorpdfstring{Calculation of $x$--$y$ Plane Parameters}{Calculation of x-y Plane Parameters}}

Now, let us find the $x$--$y$ plane parameters and consequently $\varrho$. Maxwell's electromagnetism equations state $\nabla\cdot\vec{B}=0$. Therefore, where $\rho^2=x^2+y^2$:

\begin{equation}
    \frac{\partial B_z}{\partial z}+\frac{1}{\rho}\frac{\partial}{\partial\rho}(\rho B_\rho)=0\Longrightarrow B_\rho=-\frac{1}{2}B_z'(z)\rho\Longrightarrow\frac{\partial B_\rho}{\partial \rho}=-\frac{1}{2}\frac{\partial B_z}{\partial z}
\end{equation}

Therefore, the derivative of the $z$-axis component of the field can be estimated as:

\begin{equation}
B_z(\rho,z)=B_z(0)+\frac{1}{2}B_z''(0)\left(z^2-\frac{\rho^2}{2}\right)
\end{equation}

The potential function in the $x$--$y$ plane is $V_\perp=\frac{1}{2}m\omega_\perp^2\rho^2=CB$. There are two components of the magnetic field; however, as the $z$-axis magnetic field is much larger than the radial component, the potential function can be approximated as $V=CB_z$. Therefore, 

\begin{equation}
\frac{1}{2}m\omega_\perp^2\rho^2=-\frac{1}{4}CB_z''(0)\rho^2\Longrightarrow m\omega_\perp^2=-\frac{1}{2}CB_z''(0)
\end{equation}

The negative sign of the right-hand side of this equation can be fixed with choosing $\langle C_-\rangle$. The atoms which have a positive value of $C$ will be fired out of the condensate, as there is not a sufficient trap to confine them.

Now, in order to find $a_\perp$, the estimated size of the condensate in the $x$--$y$ plane, the Schrödinger wave equation needs to be solved under the potential function we have just found. Let us assume that there are no collisions in the $x$--$y$ plane so that the scattering term in the GPE can be ignored. Therefore:

\begin{equation}
  \left( -\frac{\hbar^2}{2m} \frac{\partial^2}{\partial \rho^2} + V(\mathbf{\rho}) \right) \psi(\mathbf{\rho})=E\psi(\mathbf{\rho})
\end{equation}

Here the ground state solution to this function will be given as $\psi(\rho)=\psi_0 e^{-\rho^2/2a_\perp^2}$. Let us plug this into the equation and find $a_\perp$. After substitution and simplification:

\begin{equation}
a_\perp = \sqrt{\frac{\hbar}{m\omega_\perp}}=\left(\frac{2\hbar^2}{m|CB''(0)|}\right)^{1/4}\sim 2 \times 10^{-4}\,\mathrm{m}
\end{equation}

This gives $\frac{a_s^{{}^{87}\text{Rb}}}{a_\perp} \sim 2.5 \times 10^{-5} \ll 1$.

This indicates that the radius of the condensate cloud in the $x$--$y$ plane is much much smaller than the scattering length of ${}^{87}\text{Rb}$ atoms. Therefore:

\begin{equation}
\varrho=\frac{2 \hbar \omega_\perp a_s}{1-\zeta(1/2)\frac{a_s}{a_\perp}}\approx 2\hbar\omega_\perp a_s=100\hbar a_0\sqrt{\frac{2|CB''(0)|}{m}}
\end{equation}

\end{document}